\documentclass[sn-mathphys]{sn-jnl}
\usepackage{graphicx}%
\usepackage{multirow}%
\usepackage{amsmath,amssymb,amsfonts}%
\usepackage{amsthm}%
\usepackage{mathrsfs}%
\usepackage[title]{appendix}%
\usepackage{xcolor}%
\usepackage{textcomp}%
\usepackage{manyfoot}%
\usepackage{booktabs}%
\usepackage{algorithm}%
\usepackage{algorithmicx}%
\usepackage{algpseudocode}%
\usepackage{listings}%
\newcommand{\be}{\begin{equation}}
\newcommand{\ee}{\end{equation}}
\newcommand{\ba}{\begin{eqnarray}}
\newcommand{\ea}{\end{eqnarray}}

\theoremstyle{thmstyleone}%
\theoremstyle{thmstyletwo}%

\theoremstyle{thmstylethree}%

\begin{document}

\title[Article Title]{ B-mesons as essential probes of hot QCD matter 
}


\author*[1]{\fnm{Vinod} \sur{Chandra}}\email{vchandra@iitgn.ac.in}
\author[2]{\fnm{Santosh K.} \sur{Das}}\email{santosh@iitgoa.ac.in}
\affil*[1]{\orgdiv{Department of Physics}, \orgname{Indian Institute of Technology Gandhinagar}, \orgaddress{\city{Gandhinagar}, \postcode{382355}, \state{Gujarat}, \country{India}}}
\affil[2]{\orgdiv{School of Physical Sciences}, \orgname{Indian Institute of Technology Goa}, \orgaddress{\street{Ponda}, \city{Goa}, \postcode{403401}, \state{Goa}, \country{India}}}


\abstract{This article elucidates the pivotal role of b-mesons and bottomonium states in exploring the existence and properties of hot QCD matter (commonly known as quark-gluon-plasma (QGP) produced within the crucible heavy-ion collision experiments). Owing to the complex and confounding nature of strong interaction force the direct detection of probing the hot QCD matter is not feasible.  In light of this, investigating the dynamics of b-quarks and anti-quarks within the hot QCD medium emerges as an invaluable indirect probe. The impact of b-quarks and the mesons spans a spectrum of interesting domains regarding the physics of QCD at finite temperature, encompassing the QCD phase transition, color screening, quarkonia dissociation, heavy quark energy loss and collective flow,  anisotropic aspects, and strongly coupled nature of hot QCD medium. These aspects underscore the indispensable nature of B-mesons in the quest to create and explore the complex nature of strong interaction force through the QGP/hot QCD  matter.  In this context, we mainly focus on works related to transport studies of b-mesons in hot QCD medium,   lattice QCD, and effective field theory studies on bottomonium states, and finally, open quantum system frameworks to quarkonia to explore the properties of hot QCD medium in relativistic heavy-ion collision experiments.}




\maketitle

\section{Introduction}\label{sec1}
The study of B-mesons in heavy-ion collisions is a critical component of our quest to understand the fundamental properties of nuclear matter under extreme temperatures and energy densities. Heavy quarks~\cite{Rapp:2018qla, Cao:2018ews, Aarts:2016hap}, such as charm (c) and bottom (b), are essential for investigating and understanding the properties of hot QCD matter in relativistic heavy-ion collisions (RHIC), commonly termed as quark-gluon plasma (QGP)\cite{Gyulassy:2004zy, Shuryak:2014zxa}. Due to their larger masses, the thermalization time of heavy quarks is greater than that of the light quarks and gluons that constitute the bulk of the hot QCD matter. Thus, they traverse the hot QCD medium as non-equilibrium degrees of freedom and sense the entire spacetime evolution of the medium. Therefore, their study in heavy-ion collisions is a critical component of our quest to understand the fundamental properties of nuclear matter under extreme temperatures and energy densities, addressing the grand quest of understanding the complex nature of the strong interaction force.

The bound states of charm and bottom quarks, known as quarkonia, are integral to the exploration of the properties of the QGP medium~\cite{Djordjevic:2004nq}. Quarkonia exhibit distinct behavior in the QGP, undergoing phenomena such as sequential thermal broadening of various states and suppression, which are crucial for understanding the strongly coupled nature of the hot QCD medium. There have been extensive investigations into the in-medium properties of charmonium and bottomonium states using theoretical frameworks such as lattice QCD, incorporating the non-relativistic QCD (NRQCD) approach. By employing lattice QCD techniques, the fate of quarkonia under extreme conditions of the QGP has been explored~\cite{Ratti:2018ksb}, providing insights into their dynamics and interactions. One key aspect of studying quarkonia in the QGP is their suppression. Recent studies have been conducted regarding bottomonium suppression within an open quantum framework, where the QGP medium effectively acts as a heat bath, influencing the behavior and stability of bottomonium states. The most crucial observable in these studies has been the nuclear suppression factor, $R_{AA}$, for bottomonium, which turned out to be consistent with data from RHIC (Au-Au collisions) and LHC (Pb-Pb Collisions). For recent reviews on the usage of open quantum framework to quarkonia as a probe of the QGP, we refer the reader to~\cite{Akamatsu:2020ypb,Yao:2021lus} and for the phenomenological aspects of quarkonia production, we refer to~\cite{Lansberg:2019adr}.

In this article, we primarily concentrate on the bottom quarks and B-mesons and their pivotal role in investigating the bulk and transport properties of hot QCD matter. They provide invaluable information regarding the properties and behavior of the QGP, the phase transition from hadronic matter to the QGP, and the dynamics of heavy quarks in extreme conditions. We will now proceed into a detailed discussions of key methodologies for studying b-mesons and bottomonia within the context of a hot QCD medium, as outlined in Sections~\ref{sec2}-\ref{sec4}. The conclusive insights and future prospects will be presented in Section~\ref{sec5}.

\section{Dynamics B-mesons and b-quarks in hot QCD medium }\label{sec2}
Bottom quarks are considered as a noble probe of the hot QCD matter produced in high-energy heavy-ion collisions.  As the relaxation time
for the bottom quarks of mass $M$  at a temperature
$T$ are larger than the corresponding quantities
for light partons by a factor of $M/T(>1)$~\cite{Moore:2004tg} 
{\it i.e.} the
light quarks and the gluons get thermalized faster than the bottom 
quarks. Hence, the propagation of bottom quarks within a GQP can be conceptualized as the interplay between equilibrium and non-equilibrium degrees of freedom. The Fokker-Planck (FP) equation offers a suitable framework for describing such phenomena. To derive the Fokker-Planck equation, let us begin with the Boltzmann transport equation.

The Boltzmann transport equation that describes a non-equilibrium statistical system is given by:
\be
\left[\frac{\partial}{\partial t} 
+ \frac{\bf p}{E}.\bf{\nabla_x} 
+ {\bf F}.\bf{\nabla_p}\right]f(x,p,t)=
\left[\frac{\partial f}{\partial t}\right]_{col}
\ee
where $p$ and $E$ denote momentum and energy, ${\bf{\nabla_x}}$
(${\bf{\nabla_p}}$) are spatial (momentum space) gradient and $f(x,p,t)$
is the phase space distribution (in the present case $f$ stands for
bottom quark distribution).
The assumption of plasma uniformity and the absence of external forces
leads to
\be 
\frac{\partial f}{\partial t}=
\left[\frac{\partial f}{\partial t}\right]_{\mathrm col}
\ee
If we define $\omega(p,k)$, the rate of collisions
which change the momentum of the bottom
quark from $p$ to $p-k$, then we have~\cite{Svetitsky:1987gq}
\be
\left[\frac{\partial f}{\partial t}\right]_{col} = \int d^3k \left[ \omega(p+k,k)f(p+k) 
- \omega(p,k)f(p) \right]
\label{expeq_00}
\ee
The first term in the integrand represents a gain of probability
through collisions that knock the bottom quark
into the volume element of momentum space at $p$, and the second
the term denotes loss out of that volume element. $\omega$ is the sum of
contributions arising from the scattering of gluons, light quarks, and antiquarks.

If we expand $\omega(p+k,k)f(p+k)$ around $k$,
\be
\omega(p+k,k)f(p+k) \approx \omega(p,k)f(p) +k \frac{\partial}{\partial p} (\omega p) 
+\frac{1}{2}k_ik_j \frac{\partial^2}{\partial p_i \partial_j} (\omega p)
\label{expeq_000}
\ee

and substitute in Eq.~\ref{expeq_00}, we get: 
\be
\left[\frac{\partial f}{\partial t}\right]_{col} = 
\frac{\partial}{\partial p_i} \left[ A_i(p)f + 
\frac{\partial}{\partial p_i} \lbrack B_{ij}(p) f \rbrack\right] 
\label{expeq}
\ee
where we have defined the kernels 
\begin{eqnarray}
&& A_i = \int d^3 k \omega (p,k) k_i \nonumber\\
&&B_{ij} = \int d^3 k \omega (p,k) k_ik_j.
\end{eqnarray}
for $\mid\bf{p}\mid\rightarrow 0$,  $A_i\rightarrow \gamma p_i$ 
and $B_{ij}\rightarrow D\delta_{ij}$ where $\gamma$ and $D$ stand for
drag and diffusion coefficients respectively.
The function $\omega(p,k)$ is given by
\be
\omega(p,k)=g\int\frac{d^3q}{(2\pi)^3}f^\prime(q)v\sigma_{p,q\rightarrow p-k,q+k}
\ee
where $f^\prime$ is the phase space distribution, representing light quarks, anti-quarks, and gluons 
in the current context. $v$ represents the relative velocity between the two collision partners,
$\sigma$ denotes the cross-section and $g$ is the statistical
degeneracy. The coefficients in the first two terms of the expansion in Equation~\ref{expeq} exhibit comparable magnitudes due to greater cancellation in the averaging of $k_i$ compared to the averaging of the quadratic term $k_ik_j$. The higher powers of $k_i$'s are assumed to be relatively smaller.

With these approximations, the Boltzmann equation simplifies into a non-linear integro-differential equation, commonly referred to as the Landau kinetic equation:
\be
\frac{\partial f}{\partial t} = 
\frac{\partial}{\partial p_i} \left[ A_i(p)f + 
\frac{\partial}{\partial p_i} \lbrack B_{ij}(p) f\rbrack \right] 
\label{landaueq}
\ee
The nonlinearity arises from the presence of $f^\prime$ in $A_i$ and $B_{ij}$ through $w(p,k)$.
It emerges from the simple fact that we are investigating
collision process which involves two particles - it should,
therefore, depend on the states of the two participating particles 
and hence on the product of the two distribution functions.
Significant simplification can be attained by substituting the distribution
functions of one of the collision partners by their 
equilibrium distribution either Fermi-Dirac or Bose-Einstein
(depending on the particle type)
in the expressions of $A_i$ and $B_{ij}$. Then Eq.~\ref{landaueq} 
reduces to a linear
partial differential equation - usually referred to as the Fokker-Planck
equation. This equation describes the interaction of a particle that is out of thermal equilibrium with the particles in a thermal bath.
As mentioned earlier, the quantities $A_i$ and $B_{ij}$ are related to the usual 
drag and diffusion coefficients, and we denote them by $\gamma$ and
$D_{ij}$ respectively ({\it i.e.} these quantities can be obtained
from the expressions for $A_i$ and $B_{ij}$ by replacing the distribution
functions by their thermal counterparts):
\be
\frac{\partial f}{\partial t} = 
\frac{\partial}{\partial p_i} \left[ \gamma_i(p)f + 
\frac{\partial}{\partial p_i} \lbrack D_{ij}(p) f \rbrack \right] 
\label{FPeq}
\ee

One needs the drag and diffusion coefficient as inputs to solve the Fokker-Planck equation along with the initial distribution. The drag coefficient can be defined as the thermal average
of the momentum transfer weighted by the square of the invariant transition amplitude. For the elastic two-body collisional process $HQ(P)+l(Q)\rightarrow HQ(P^{'})+l(Q^{'})$, where $l$ represents constituent particles in the thermal medium (quarks, antiquarks, and gluons) and $P, Q$ are the four-momentum of the bottom quark and constituent particle before the collision, the heavy quark drag and momentum diffusion can be described as, 
 \begin{align}\label{15}
    A_i=&\frac{1}{2E_p}\int{\frac{d^3{\bf q}}{(2\pi)^32E_q}}\int{\frac{d^3{\bf q}'}{(2\pi)^32E_{q'}}}\int{\frac{d^3{\bf p}'}{(2\pi)^32E_{p'}}}\frac{1}{\gamma_{HQ}}\nonumber\\ &\times\sum|\mathcal{M}_{2\rightarrow 2}|^2(2\pi)^4\delta^4 (P+Q-P'-Q') f_{k}({\bf{q}})\nonumber\\ &\times\Big(1+a_k f_{k}({\bf{q'}})\Big)\Big[({\bf p}-{\bf p}')_i\Big]\nonumber\\
	&=\langle\langle({\bf p}-{\bf p}')_i\rangle\rangle,
\end{align}
  and 
\begin{align}\label{16}
    B_{ij}=&\frac{1}{2E_p}\int{\frac{d^3{\bf q}}{(2\pi)^32E_q}}\int{\frac{d^3{\bf q}'}{(2\pi)^32E_{q'}}}\int{\frac{d^3{\bf p}'}{(2\pi)^32E_{p'}}}\frac{1}{\gamma_{HQ}}\nonumber\\ &\times\sum|\mathcal{M}_{2\rightarrow 2}|^2(2\pi)^4\delta^4 (P+Q-P'-Q') f_{k}({\bf{q}})\nonumber\\ &\times\Big(1+a_k f_{k}({\bf{q'}})\Big)\frac{1}{2}\Big[({\bf p}-{\bf p}')_i({\bf p}-{\bf p}')_j\Big]\nonumber\\
	&=\frac{1}{2}\langle\langle({\bf p}-{\bf p}')_i({\bf p}-{\bf p}')_j\rangle\rangle,
\end{align} 

respectively. It is important to note that the delta function guarantees energy-momentum conservation in the system. and $f_{k}$ is the equilibrium phase space distribution for the light quarks, antiquarks, and gluons. Here, $\gamma_{HQ}$ denotes the statistical degeneracy factor of the heavy quark, and   $|\mathcal{M}_{2\rightarrow 2}|$ represents the matrix element for the two-body elastic collisions of the bottom with light quarks, antiquarks, and gluons~\cite{Svetitsky:1987gq}. It is crucial to highlight that the drag coefficient for the bottom quark represents the thermal average of the momentum transfer, while the momentum diffusion captures the square of the momentum transfer resulting from the interaction. As $A_{i}$ depends on the momentum, we have the following decomposition of the bottom quark drag,
\begin{align}\label{17}
&A_{i}=p_{i}A(p^2), &&A=\langle\langle 1 \rangle\rangle - \frac{\langle\langle {\bf{p.p'} \rangle\rangle}}{p^2}, 
\end{align}
where $p^2=|{\bf p}|^2$ and $A$ is the drag coefficient of the heavy quark. Similarly, the momentum diffusion $B_{ij}$ can be decomposed in terms of longitudinal and transverse components as follows, 
\begin{align}\label{18}
&B_{i,j} = \left(\delta_{ij}-\frac{p_ip_j}{p^2}\right) B_0(p^2)+\frac{p_ip_j}{p^2}B_1(p^2),
\end{align}
with the transverse and longitudinal diffusion coefficients, they respectively assume the following forms:
\begin{align}\label{19}
&B_{0}= \frac{1}{4}\left[\langle\langle p'^{2} \rangle\rangle-\frac{\langle\langle ({\bf{p'.p}})^2\rangle\rangle}{p^2} \right],\\ 
&B_{1}= \frac{1}{2}\left[\frac{\langle\langle ({\bf{p'.p})}^2\rangle\rangle}{p^2} -2\langle\langle ({\bf{p'.p})}\rangle\rangle +p^2 \langle\langle 1 \rangle\rangle\right]\label{19.1}.
 \end{align}

 The standard approach to HQ dynamics in the
QGP is to follow their evolution using a
 Fokker-Plank equation solved stochastically by the
Langevin equations. The relativistic Langevin equations of motion governing the evolution of the position and momentum of the heavy quarks can be expressed in the following form:~\cite{Das:2015ana}:

\begin{eqnarray}
 dx_i=\frac{p_i}{E×}dt, \nonumber \\
 dp_i=-A p_i dt+C_{ij}\rho_j\sqrt{dt}
 \label{lv1}
\end{eqnarray}
where $dx_i$ and $dp_i$ are the changes of the coordinate and momentum in each time step $dt$.
$A$ and $C_{ij}$ are the  drag coefficient and the
covariance matrix in terms of independent Gaussian-normal 
distributed random variables $\rho$,$P(\rho)=(2\pi)^{-3/2}e^{-\rho^2/2}$, which obey 
the relations $<\rho_i \rho_j>=\delta_{ij}$ and $<\rho_i>=0$, respectively. 
The covariance matrix is related to the diffusion tensor, 
\begin{eqnarray}
C_{ij}=\sqrt{2B_0}P_{ij}^{\perp}+\sqrt{2B_1}P_{ij}^{\parallel}, 
\label{cmmm}
\end{eqnarray}
where $P_{ij}^{\perp}= \delta_{ij}-p_i p_j/p^2$ and $P_{ij}^{\parallel}=p_i p_j/p^2$ 
represent the transverse and longitudinal projector operators, respectively.
With the assumption, $B_0=B_1=D$, 
Eq~(\ref{cmmm}) becomes $C_{ij}=\sqrt{2D(p)} \delta_{ij}$. This assumption is strictly valid only for $p\rightarrow 0$; however, it is commonly employed at finite $p$ when applied to the dynamics of bottom (also charm quark) quarks in hot QCD matter.
 
The Fokker-Planck or Langevin equation results can be utilized to calculate experimental observables, such as the nuclear suppression factor, $R_{AA}$, and elliptic flow, $v_2$.
The nuclear suppression factor, $R_{AA}$, can be evaluated using the initial 
charm and bottom quark distributions at initial time $t=\tau_i$ 
and final time $t=\tau_f$ at the freeze-out temperature  as $R_{AA}(p)=\frac{f(p,\tau_f)}{f(p,\tau_i)}$. 
Along with $R_{AA}$, the anisotropic momentum distribution induced by 
the spatial anisotropy of the bulk medium can be calculated and defined as
\be
 v_2=\left\langle  \frac{p_x^2 -p_y^2}{p_x^2+p_y^2}\right\rangle \ , \qquad \qquad
\ee
\label{eq5}
which measures the momentum space anisotropy.

\subsection{Boltzmann vs Fokker-Planck dynamics}\label{subsec1}
The propagation of heavy quarks, mainly charm and bottom, in QGP has been quite often treated within the framework of the Fokker-Planck equation. This is primarily because their motion can be likened to Brownian motion, attributed to their perturbative interaction and large mass. This characteristic typically results in collisions that are sufficiently forward-peaked and/or entail small momentum transfers.
Under these constraints, it is recognized that the Boltzmann transport equation also simplifies to Fokker-Planck dynamics. This reduction represents a substantial simplification of in-medium dynamics.
 This approach has been extensively utilized in the literature for the computation of experimentally observed nuclear suppression factor ($R_{AA}$) and elliptic flow ($v_2$).

Indeed, the initial motivation for adopting a Fokker-Planck approach was primarily linked to the assumption that heavy quarks, both charm and bottom quarks, experience relatively small momentum transfers.  
However, the experimentally observed suppression factors ($R_{AA}$) and elliptic flows ($v_2$) for light and charmed hadrons in the QGP indicate that the momentum transfer for heavy quarks may not be sufficiently small. Hence, investigating the effects of the approximations inherent in the Fokker-Planck equation through a direct comparison with the full collisional integral within the Boltzmann transport equation framework would be valuable. This analysis could help determine whether heavy quarks indeed exhibit Brownian motion in the QGP.

The momentum evolution of heavy quarks is explored using both the Langevin and Boltzmann approaches, with the initial distribution of charm and bottom quarks assumed to be a delta distribution at p=10 GeV.
The Debye mass is taken to be $m_D=gT$ to regularize the divergence arising from the t-channel process.
The heavy quark drag and diffusion coefficients with the
gluons in the bulk have been computed within the framework of perturbative Quantum Chromodynamics (pQCD) to serve as input in the Langevin equation. 
Numerical solutions for the Boltzmann equation are obtained by discretizing space into a three-dimensional lattice and employing the test particle method to sample distribution functions. The collision integral is addressed through a stochastic implementation of the collision probability, where the heavy quark-gluon cross-section is treated as an input, computed within the framework of pQCD. For further details, please refer to reference ~\cite{Das:2013kea}.

\begin{figure*}
 \centering
 {\includegraphics[scale=0.23]{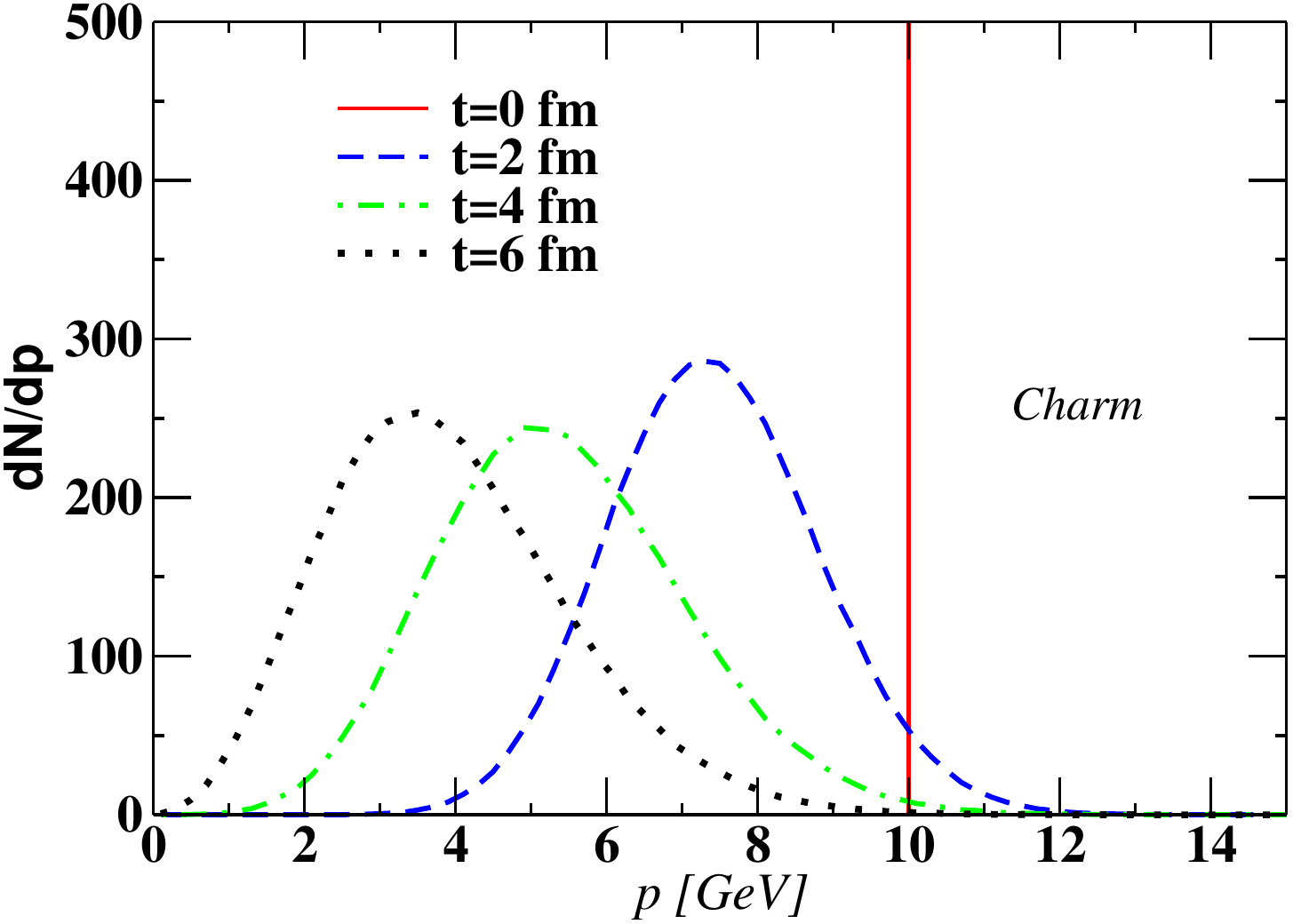}}
 \hspace{1 cm}
 {\includegraphics[scale=0.23]{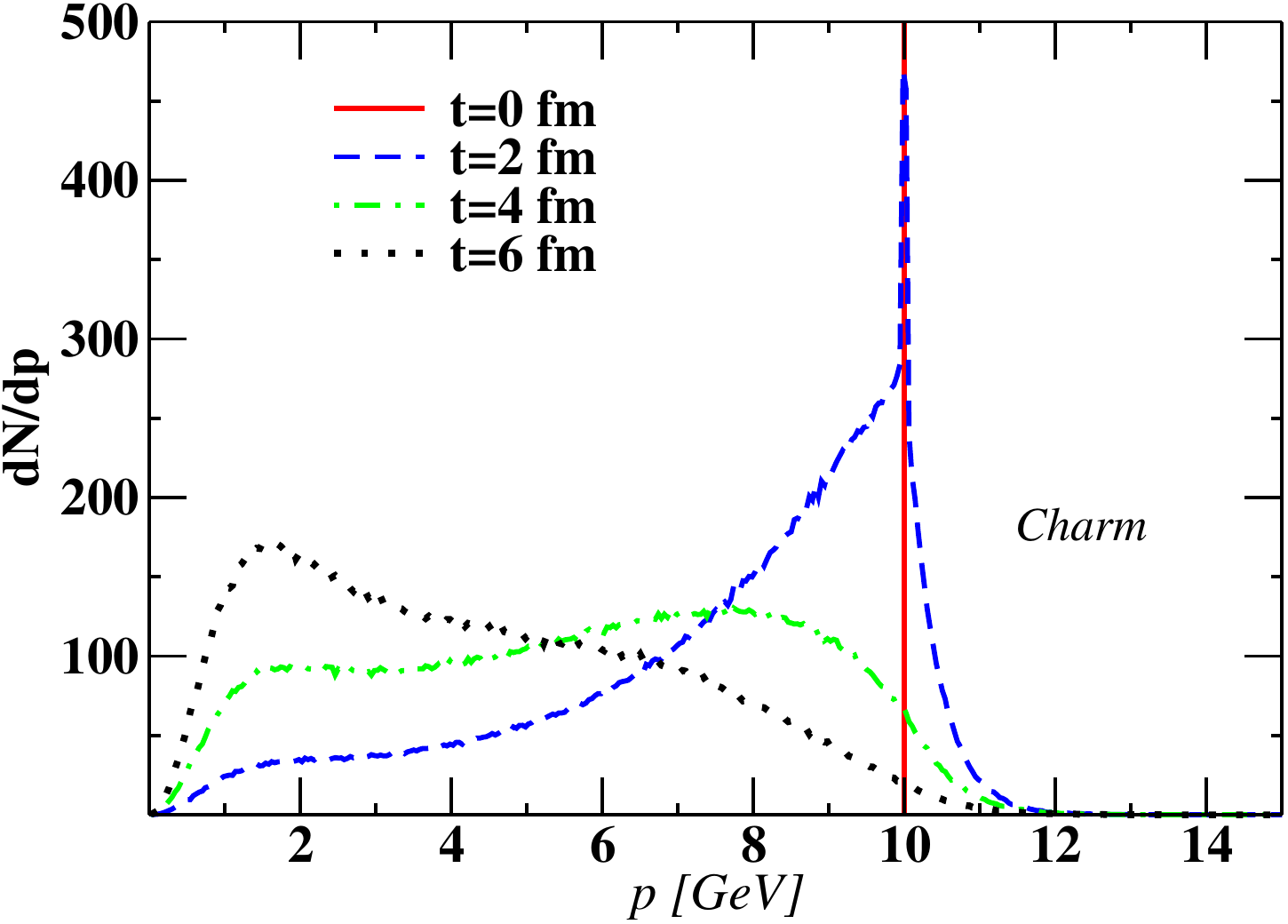}}

\caption{Left panel: Charm quark momentum evolution obtained within Langevin dynamical considering the initial momentum distribution of the charm quarks as a delta distribution at p=10 GeV. Right panel: Charm quark momentum evolution obtained within the Boltzmann equation considering the initial momentum distribution of the charm quarks as a delta distribution at p=10 GeV. Figures adapted from Ref. ~\cite{Das:2013kea}.}
\label{Tcc}
\end{figure*}

\begin{figure*}
 \centering
 {\includegraphics[scale=0.23]{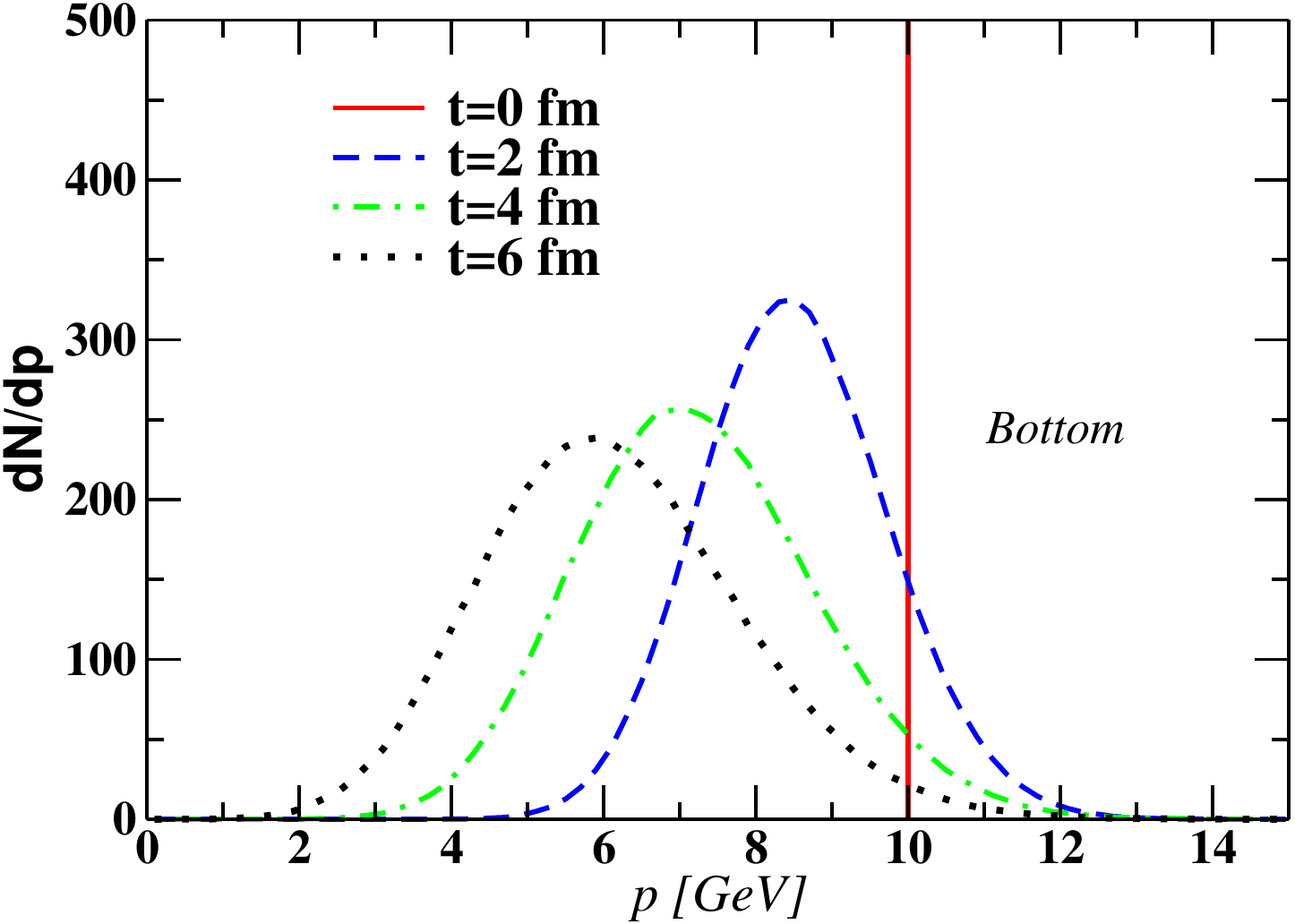}}
 \hspace{1 cm}
 {\includegraphics[scale=0.23]{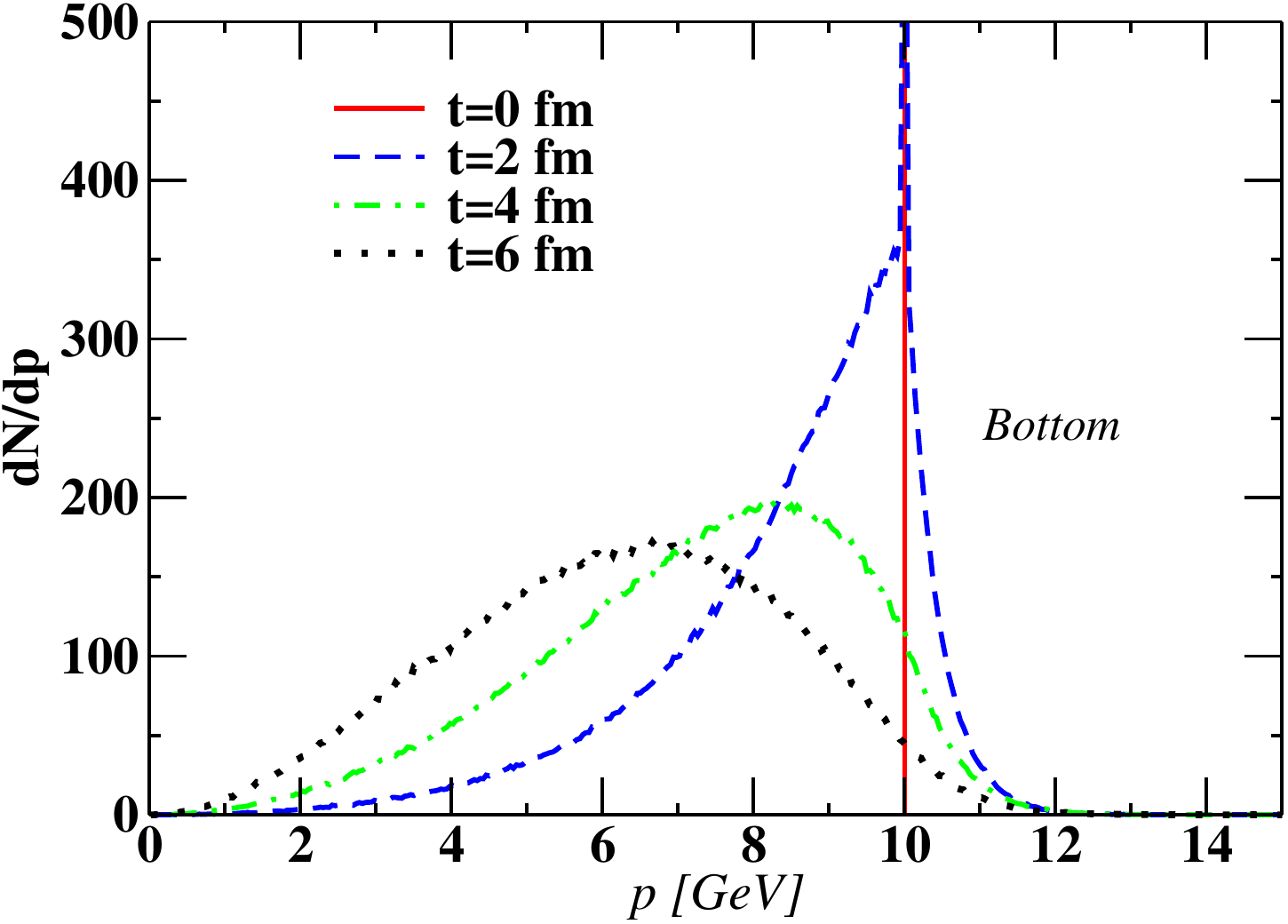}}

\caption{Left panel: Bottom quark momentum evolution obtained within Langevin equation considering the initial momentum distribution of the bottom quarks as a delta distribution at p=10 GeV. Right panel: Bottom quark momentum evolution obtained within the Boltzmann equation considering the initial momentum distribution of the bottom quarks as a delta distribution at p=10 GeV. Figures adapted from Ref. ~\cite{Das:2013kea}.}
\label{Tbb}
\end{figure*}

 The objective is to compare the time evolution of charm and bottom quarks, originating from the same initial momentum distribution within a bulk at a temperature of T=0.4 GeV, using both the Langevin and Boltzmann equations. The momentum evolution of the charm quarks is displayed in Fig.\ref{Tcc} (left panel) obtained within the Langevin dynamics. Fig.\ref{Tbb} (left panel) obtained within the Langevin dynamics displays the bottom quark momentum evolution.  The observation reveals that both charm and bottom quarks exhibit Gaussian distributions, as anticipated by the construction of the model. This present calculation indicates that the equilibrium distribution can be achieved at the end of the evolution by implementing the Fluctuation-dissipation theorem (FDT). 
As is well-known, the Langevin dynamics consists of a shift of the
average momenta accompanied by fluctuations around this average, allowing the heavy quark the possibility of gaining energy. This is evident from the tail of the momentum distribution that overshoots the initial
momentum p = 10GeV at t = 2 fm/c, the black solid line in Fig.\ref{Tcc} (left panel) and Fig.\ref{Tbb} (left panel).

In Fig. \ref{Tcc} (right panel), the momentum distribution for charm quarks within the Boltzmann equation is depicted. There is a distinctly different evolution of the particle momentum, characterized by a non-Gaussian shape. By t=2vfm/c, the distribution displays a broader spread in momentum, with a notable contribution from processes where charm quarks can gain energy. Additionally, there is a long tail at low momenta, indicating a probability of losing a significant amount of energy. Overall, the global shape deviates significantly from a Gaussian form. This observation suggests that the evolution of the charm quark at a temperature T=0.4 GeV is not Brownian in nature. In Fig.\ref{Tbb} (right panel) we present the momentum distribution for the bottom quark obtained within the Boltzmann equation. For the bottom quarks, shown in Fig.\ref{Tbb} (right panel), the momentum evolution gives a much better agreement between the Boltzmann and the Langevin evolution because $M_b/T\simeq $ 10. This suggests that the Langevin approach serves as a good approximation for the bottom quark, while for the charm quark, there can be significant deviations~\cite{Das:2013kea}.

 To compute the $R_{AA}$ and $v_2$ of B-meson, bottom quarks are distributed in momentum space according to the Fixed Order + Next-to-Leading Log (FONLL)~\cite{Cacciari:2005rk} calculation, which describes the B-mesons spectra in proton-proton collisions. The solution of the Langevin equation can be convoluted with the fragmentation functions of the bottom quarks at the quark-hadron transition temperature $T_c$
to obtain the momentum distribution of the B mesons. One of the prime goals of all phenomenological studies of open heavy-flavor observables is to extract the heavy quark spatial diffusion coefficient, $D_x$, which quantifies the interaction of heavy quarks with the bulk medium. $D_x$ is also directly related to the heavy quark thermalization time and can be evaluated using lattice QCD (lQCD). Transport models are quite successful in describing the experimentally measured observables, the nuclear modification factor, $R_{AA}$,  and elliptic flow, $v_2$, of the B-meson.
Drag and momentum diffusion coefficients are inputs in the transport models (cross sections are the inputs in the Boltzmann equation), which contain the microscopic details mentioning how the heavy quarks interact with the hot QGP medium. Bottom quark drag and diffusion coefficients can be computed within the framework of pQCD.  However, pQCD does not work to describe the measured heavy quark $R_{AA}$ and $v_2$ 
at RHIC and LHC~\cite{Das:2015ana}. At low $p_T$ and low T, the non-perturbative effect plays a crucial role in simultaneously describing both the  $R_{AA}$ and $v_2$ of heavy quark. Non-perturbative effects can be taken into account through the quasi-particle model (QPM)~\cite{Das:2015ana}. 
Then, they compare their results with the experimental data. The value of the transport coefficients with which they can describe the 
experimental data, they relate it with the spatial diffusion coefficient, $D_x$.
The spatial  diffusion coefficient, $D_x$, 
can be calculated in the static limit ($p\rightarrow 0$) from the drag coefficient $D_x=T/M F$, where $T$ is the temperature of the thermal bath, $M$ is the mass of the heavy quark and $F$ is the drag coefficient. The standard quantification of the space
diffusion coefficient is done in terms of a dimensionless quantity $2\pi T D_x$ ~\cite{He:2012df} which is independent of mass.  One can estimate the charm quark thermalization time as $ \tau_{th} \equiv \Gamma^{-1}(p\rightarrow 0)$~\cite{Scardina:2017ipo}:

\begin{figure*}
 \centering
 {\includegraphics[scale=0.35]{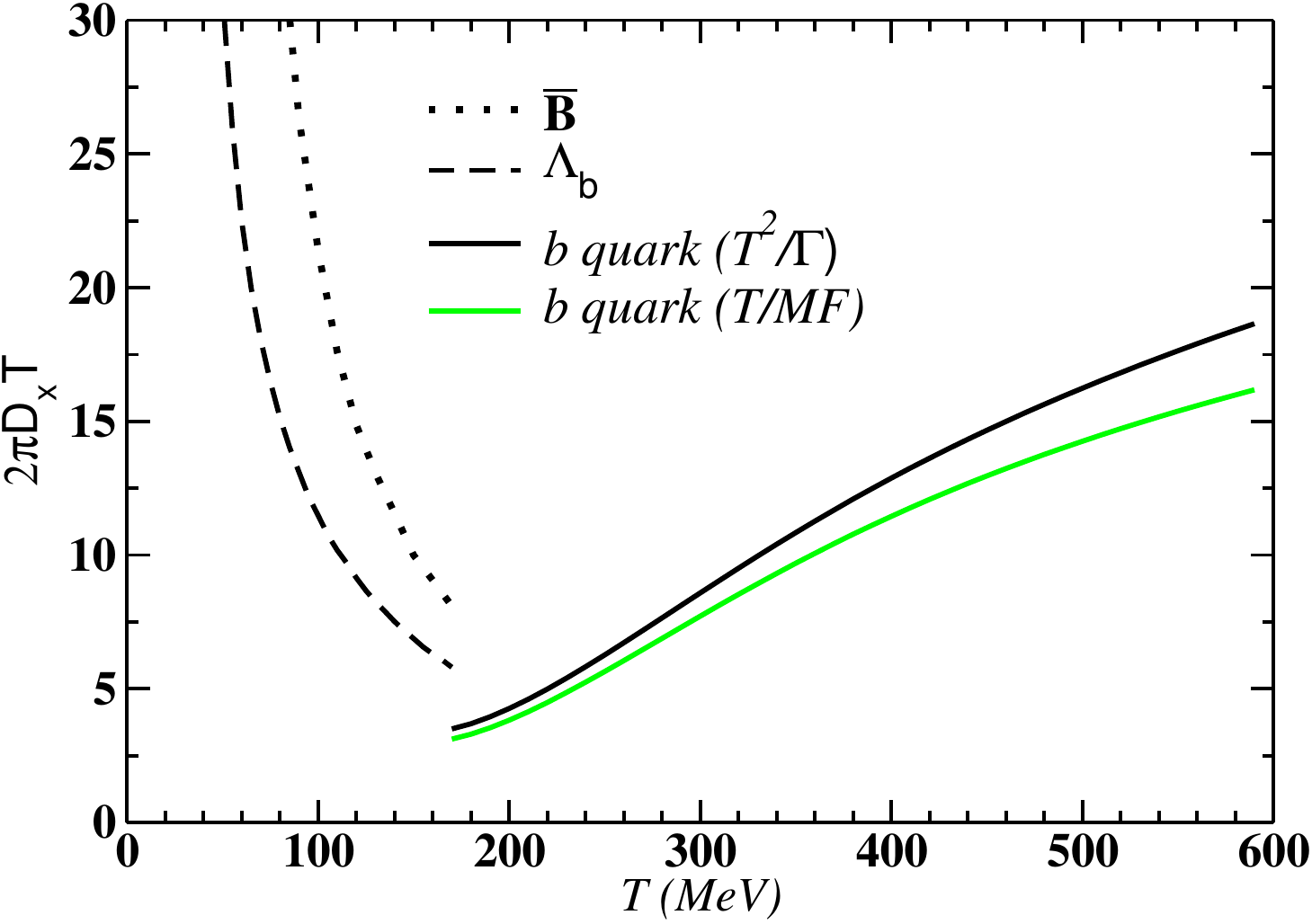}}
 \hspace{1 cm}

\caption{: Spatial diffusion coefficients $D_x$ (in units of the thermal wavelength, 1/(2$\pi$T )) as a function of temperature (T) for b quark, $\bar{B}$ and $\Lambda_b$. Figure adapted from Ref. ~\cite{Das:2016llg}.}
\label{Dbb}
\end{figure*}

\begin{equation}
\tau_{th} = \frac{M D_x}{T}  = \frac{M}{2\pi T^2} (2\pi T D_x) \cong 1.8 \, \frac{2\pi T D_x}{(T/T_c)^2} \,\, \rm fm/c,
\label{eq:th}
\end{equation}

where $T_c$ is the quark-hadron transition temperature.  The $D_x$ can also computed from the diffusion coefficient ($\Gamma$) through the fluctuation-dissipation theorem (FDT), $\Gamma= M F T$.

In Fig.\ref{Dbb}, the variation of $D_x$ is shown as a function of temperature for b quarks (high
temperature), $\bar{B}$ and  $\Lambda_b$ (low temperature).  We notice
that the $\bar{B}$ and $\Lambda_b$ diffusion coefficient also support a continuous evolution with a minimum around $T_c$ like the D meson case. This indicates the strength of the interaction is the maximum near the quark-hadron transition temperature. The diffusion coefficient of the $\bar{B}$ and $\Lambda_b$ in the hadronic matter is shown in  Fig.\ref{Dbb} computed within the effective field theory framework. For
further details, please refer to reference~\cite{Tolos:2016slr}.

\section{B-mesons at finite temperature in the lattice}\label{sec3}
The in-medium characteristics of bottomonium have been explored through lattice QCD investigations, as documented in previous studies~\cite{Aarts:2010ek,Aarts:2013kaa,Aarts:2014cda,Kim:2014iga,Kim:2018yhk}. These investigations primarily concentrate on discerning the modifications occurring in the ground states of S- and P-wave quarkonia within the medium. However, a notable challenge in these lattice studies arises from the use of point meson operators, which exhibit limited overlap with quarkonia wavefunctions, particularly those associated with excited states. Consequently, the correlators tend to be predominantly influenced by the vacuum continuum portions of the spectral function, posing a significant obstacle in accurately computing in-medium bottomonium properties~\cite{Burnier:2015tda,Mocsy:2007yj}.

Efforts have been made to address this limitation by employing extended meson operators, known for their improved compatibility with S- and P-state bottomonia~\cite{Larsen:2019bwy}. However, further refinement is needed to delve into the in-medium properties of excited bottomonium states, such as 3S and 2P-states. A step in this direction has been taken by introducing extended meson operators within the lattice NRQCD framework~\cite{Larsen:2019zqv}, revealing evidence of sequential in-medium modifications aligned with the sizes of the bottomonium states.

The investigation into the thermal broadening of quarkonia states within a lattice framework remains an active research area. More nuanced and refined results in this domain are anticipated, especially considering the temperature regimes accessible at RHIC and LHC energies. This ongoing work is crucial for advancing our understanding of the behavior of bottomonium in a hot and dense medium.

\section{Bottomonium suppression within open quantum system framework}\label{sec4}
Quankonia suppression was proposed as a potential indicator of the QGP format long ago by Matsui and Satz~\cite{Matsui:1986dk}. Since then quakonia suppression has been an integral part of heavy-ion physics investigations in experiments at RHIC and LHC. The idea of Matsui was primarily based on the screening of chromoelectric fields at a distance inverse of the Debye mass ($m_D\sim g(T) T$, where $g$ is QCD coupling constant and $T$ is the temperature of the system. In this picture, whenever the binding energy of a given quakonia is overcome by the temperature the said quakonia dissociates. This mechanism has been revised by the first principle QCD computations at finite temperature that computes the heavy quark-antiquark static potential in the QGP medium~\cite{Laine:2006ns}. The potential turned out to the complex. The imaginary part leads to the thermal width and it is this part that dominates the in-medium heavy quarkonia suppression rather than the screening effects. Employing this potential in non-relativistic QCD effective theory \cite{Brambilla:2013dpa, Brambilla:2017zei} framework, it is shown that the in-medium quantum evolution of heavy quarkonium depends on its momentum diffusion coefficient and its dispersive counterpart. Both the transport coefficients could be defined in terms of nonperturbative correlators of chromoelectric fields.

On the other hand,  Non-relativistic Effective Field Theories (NREFTs) serve the purpose of appropriately defining the heavy quark potential to enable a way for the systematic calculation of quarkonia properties in the hot QCD medium. This framework relies on the separation of various scales that are characteristic of NR bond states (both at zero as well as finite temperatures). Heavy quark mass, $m$, the inverse of the Bohr radius, $1/r_0\sim m v$, and the binding energy, $e\sim m v^2$ ($v<<1$, being the heavy quark velocity) are the energy scales at zero temperature. At finite temperature,  more scales enter in the picture, for instance, the temperature T and, at weak coupling, the Debye mass $m_D$. Both the cases Nevertheless, at leading order in $v$ the equation of motion is still a Schrödinger equation, which describes the real-time evolution of the $Q\bar{Q}$  pair in the medium. The effective field theory (EFT)  framework that is obtained by integrating out degrees of freedom associated with the scale $m$ is Non-Relativistic QCD (NRQCD)~\cite{Caswell:1985ui, Bodwin:1994jh}. The EFT that integrates out gluons with momentum or energy scaling like the inverse of the Bohr radius is the potential NRQCD (pNRQCD)~\cite{Pineda:1997bj,Brambilla:1999xf, Brambilla:2004jw}.  At the leading order in heavy quark velocity, $v$, the equation of motion of pNRQCD is the Schro"dinger equation for an NR bound state. Note that pNRQCD provides an unambiguous field theoretical definition of the potential (encoding the contributions coming from modes with energy and momentum above the scale of the binding energy). This framework includes systematically higher order corrections to leading of Schr{o}'dinger equation. The first corrections in the weak coupling regime are carried out by chromoelectric dipole terms.

The essential processes regarding quarkoina physics in the QGP, {viz.}, Quarkonium scattering in the medium, its dissociation into an unbound color octet $Q\bar{Q}$ pair, and the inverse processes of $Q\bar{Q}$ pair generation need a framework that describes the quarkonium non-equilibrium evolution in the said medium. This calls for the open quantum system framework (OQS) where the background QGP medium plays the role of the environment with temperature scale T. The OQS framework in this context was proposed in a seminal work by Akamatsu~\cite{Akamatsu:2014qsa}.  
 
In the context of bottomonium, the suppression has been investigated in an OQS framework in~\cite{Brambilla:2020qwo}. In this work,  the Lindblad equation has been derived in the framework of pNRQCD accounting for the quantum and non-Abelian nature of the quarkonia system.  The bottomonium nuclear suppression factor computed within the framework taking lattice QCD inputs was shown to agree well with the LHC data for 5.02 TeV Pb-Pb collisions. In a very recent work~\cite{Strickland:2023nfm}, NLO correction has been considered to above mentioned OQS framework within pNRQCD, and predictions on bottomonium suppression were found to be in good agreement with LHC data but some disagreements were found with RHIC data.

These investigations have gained momentum and we hope for a more refined understanding regarding bottomonium production and suppression in the hot QCD medium shortly. It is to be noted that the usage of the OQS framework in high-energy physics extends beyond the confines of the heavy-quarkonia transport. The OQS framework has been employed in diverse areas such as dark matter formation~\cite{Binder:2020efn}, deeply inelastic reactions~\cite{Braaten:2016sja}, and studies related to inflation~\cite{Boyanovsky:2015tba,Boyanovsky:2015jen} etc.,  demonstrating its applicability across a spectrum of scientific investigations.

\section{Conclusions and outlook}\label{sec5}
The study of the properties of hot QCD matter is a field of
contemporary interest and the heavy quarks (HQ) namely, charm and bottom quarks
play crucial roles in this endeavor. Heavy flavor particles are not anticipated to undergo complete thermalization. As non-equilibrium probes, they retain a memory of their interaction history, providing a gauge of their interaction strength with the medium.  Unlike charm quarks, bottom quarks exhibit behavior akin to Brownian motion, and the Langevin approach provides a suitable approximation for their dynamics. However, the momentum evolution of charm quarks is not Brownian in nature at the temperatures anticipated in experiments conducted at the Relativistic Heavy Ion Collider (RHIC) and the Large Hadron Collider (LHC). Additionally, it has been suggested that the heavy quark diffusion with an infinite-mass limit may not be a suitable approximation for the charm quark \cite{Pandey:2023dzz}. However, for the dynamics of the bottom quark in QCD matter, the infinite-mass limit remains a reasonable approximation. This emphasizes the significance of bottom quarks as a probe of the hot QCD matter. They possess the potential to establish a connection between the phenomenology constrained by experimental data and lQCD, offering insights into the transport properties of the hot QCD matter.

The presence of the hadronic phase is unavoidable in high-energy heavy-ion collisions. Following the QGP phase, the bottom quark undergoes hadronization, resulting in the formation of B-mesons. Subsequently, these B-mesons interact with the surrounding hadronic matter, which comprises particles such as pions, kaons, etas, and nucleons. Rescattering within the hadronic medium can have a significant impact on observables related to B-mesons, particularly their elliptic flow. The study of these effects constitutes a vibrant and active field of research~\cite{,Das:2011vba,Torres-Rincon:2014ffa}. Furthermore, the ratio of heavy baryons to mesons ($\Lambda_c/D$ and $\Lambda_b/B$) is of fundamental importance in gaining insights into in-medium hadronization processes~\cite{Das:2016llg}. The ratio $\Lambda_b/B$ can serve as a valuable tool for disentangling different hadronization mechanisms. This is one of the primary focuses of the ongoing experimental efforts at both RHIC and LHC colliding energies, aimed at understanding the mechanisms of hadronization. The B-meson, acting as a non-equilibrium probe and being produced early in the collision process, offers an excellent avenue for probing the early stages of heavy-ion collisions. This is a rapidly emerging field of research, with ongoing investigations aimed at gaining a deeper understanding of the dynamics and properties of the early stages of these high-energy collisions~\cite{Das:2023zna}.

The OQS framework, incorporating inputs from effective field theory and lattice QCD (lQCD) within the context of bottomonium, proves to be a promising approach for examining its behavior in the hot QCD medium. The findings related to quarkonia suppression align consistently with data obtained from RHIC and LHC experiments. Anticipated to undergo essential refinements, this methodology is poised to yield more reliable results in the near future.

\section*{Acknowledgement} We would like to acknowledge the Science and Engineering Research Board (SERB) for  Core Research Grant (CRG) No. CRG/2020/002320.
\\
\\

\noindent {\bf Data availability statement}: No data associated with the manuscript.

\bibliography{sn-bibliography}

\end{document}